\documentclass{PoS}
\usepackage{epsfig}
\usepackage{graphicx}

\title{Two-Flavor Chiral Perturbation Theory for Hyperons}

\ShortTitle{Two-Flavor Chiral Perturbation Theory for Hyperons}

\author{\speaker{B.~C.~Tiburzi}\\
        Maryland Center for Fundamental Physics\\
        Department of Physics\\
        University of Maryland\\
        College Park, MD 20742-4111, USA\\
        E-mail: \email{bctiburz@umd.edu}
        }

\abstract{
The three-flavor chiral expansion for octet baryons has well-known problems with convergence. 
We show that this three-flavor chiral expansion can be reorganized into a two-flavor expansion
thereby eliminating large kaon and eta loop contributions. 
Issues of the underlying formulation are addressed by considering the effect 
of strangeness changing thresholds on hyperon masses. 
While the spin-3/2 hyperon resonances are considerably more sensitive
to these thresholds compared to the spin-1/2 hyperons, 
we demonstrate that in both cases the essential physics can be captured in 
the two-flavor effective theory by terms that are analytic in the pion mass squared,
but non-analytic in the strange quark mass. 
Using the two-flavor theory of hyperons, 
baryon masses and axial charges are investigated.
Loop contributions in the two-flavor theory appear to be perturbatively under control. 
A natural application for our development is to study the pion mass dependence of lattice QCD data on hyperon properties. 
}

\FullConference{6th International Workshop on Chiral Dynamics, CD09\\
		July 6-10, 2009\\
		Bern, Switzerland}
		
\begin{document}

\section{Introduction and Motivation}

Before the theory of QCD was established, 
patterns observed in the spectrum of low-lying hadrons shed light on underlying symmetries. 
The lowest-lying mesons and baryons appear to assemble themselves into 
SU(3) 
multiplets:
an octet of pseudoscalar mesons, 
an octet of spin-1/2 baryons, 
and a decuplet of spin-3/2 baryons. 
The lightness of the octet of pseudoscalar mesons suggests that they are Goldstone bosons emerging from 
spontaneous chiral symmetry breaking:
SU(3)${}_L \times$ SU(3)${}_R \to$ SU(3)${}_V$. 
Their non-vanishing masses would arise from explicit chiral symmetry breaking introduced by 
quark masses in the QCD Lagrangian. 
The remaining hadrons would naturally be grouped into multiplets of the unbroken SU(3)${}_V$ symmetry. 
This picture is rigorous in the limit of small quark masses, 
$m_u$, $m_d$, $m_s \ll \Lambda_{QCD}$, 
for which chiral perturbation theory 
(ChPT) 
becomes an effective description of the low-energy 
dynamics of pseudoscalar mesons and baryons.

In nature, 
however, 
things are less pristine:
the size of the strange quark mass is not considerably smaller than the QCD scale;
and, 
for some quantities, 
SU(3) 
appears to be badly broken. 
We review the treatment of the octet baryons within 
SU(3) 
ChPT. 
We focus on the strengths and weaknesses of such an approach, 
ultimately arguing in favor of a separate
SU(2) 
treatment for nucleons and the various hyperons. 
Recently there has been work on strange hadrons in SU(2) ChPT%
~\cite{Roessl:1999iu,Frink:2002ht,Beane:2003yx,Allton:2008pn,Tiburzi:2008bk,Flynn:2008tg,Jiang:2009sf,Mai:2009ce}.
Such theories emerge naturally from reordering the 
SU(3) 
expansion,
as we demonstrate for the baryon masses. 
Particular attention is paid to the effect of strangeness changing thresholds. 
We find that such virtual processes can be well described by 
terms analytic in the pion mass squared, but non-analytic in the strange quark mass.
Hyperon axial charges are also considered in this framework. 
Prospects for the two-flavor ChPT of hyperons
are summarized at the end.

\section{Baryon Masses}

At the leading order (LO) in ChPT, 
pseudoscalar mesons 
(the pion, kaon, and eta) 
satisfy the 
SU(3) 
mass relation
\begin{equation}
\Delta_{GMO} (m_\phi^2)
= 
\frac{4}{3} m_K^2 - m_\eta^2 - \frac{1}{3} m_\pi^2
=
0
,\end{equation}
originally due to Gell-Mann and Okubo. 
Experimentally this LO relation is fairly well satisfied.
Using the neutral meson masses, and 
dividing by the average mass of the octet, 
the relation is satisfied at the level of $14.5 \%$.
This small excess can be attributed to next-to-leading order (NLO)
contributions in the SU(3) chiral expansion; and, from the excess, 
the value of a certain combination of low-energy constants can be determined
and is of natural size.

For the octet baryons, there is additionally a Gell-Mann--Okubo relation 
between their masses. 
In heavy baryon ChPT (HBChPT), 
this relation arises because there are four isospin symmetric masses,
but only three LO chiral symmetry breaking operators:
$< \overline{B} \{ m_q, B \} >$, 
$< \overline{B} [ m_q, B]>$, 
and 
$< \overline{B} B >\, <m_q>$. 
One consequently has the relation
\begin{equation}
M_{GMO} 
= 
M_\Lambda 
+ 
\frac{1}{3} M_\Sigma
- 
\frac{2}{3} M_N
- 
\frac{2}{3} M_\Xi
=
0
.\end{equation}
Using the neutral baryon masses divided by the weighted average of the octet baryon mass,
this relation is satisfied to a remarkable level experimentally: $0.90 \%$.  
Furthermore, 
the NLO order contribution to $M_{GMO}$ can be determined in HBChPT
in terms of reasonably well-known axial couplings
\begin{equation} \label{eq:MGMO}
M_{GMO}
=
\frac{4}{3 \Lambda_\chi^2}
\left[
\pi ( D^2 - 3 F^2) \Delta_{GMO} (m_\phi^3)
- 
\frac{1}{6} C^2 \Delta_{GMO} \Big( \mathcal{F} (m_\phi, \Delta, \mu ) \Big)
\right]
,\end{equation}
where $\mathcal{F}$ is the non-analytic function arising from the sunset diagram,
\begin{equation}
\mathcal{F}
(m,\delta,\mu)
=
(m^2 - \delta^2) 
\left[ 
\sqrt{\delta^2 - m^2} 
\log 
\left(
\frac{\delta - \sqrt{\delta^2 - m^2 + i \varepsilon}}{\delta + \sqrt{\delta^2 - m^2 + i \varepsilon}}
\right)
- 
\delta \log 
\frac{m^2}{\mu^2}
\right] 
- 
\frac{1}{2} \delta m^2 
\log 
\frac{m^2}{\mu^2}
,\end{equation} 
and 
$\Delta$ 
is the average splitting between the decuplet and octet baryons. 
By SU(3) symmetry,
the $\mu$-dependence in Eq.~(\ref{eq:MGMO}) is only superficial.  
For various estimates of the axial couplings collected in Table~\ref{t:GMO}, 
we arrive at the right size for 
$M_{GMO}$, 
with the difference presumably due to NNLO terms.

%
  \begin{table}
  \begin{center}
     \caption{Estimates of the Gell-Mann--Okubo baryon mass relation from NLO SU(3) HBChPT.  }
    \smallskip
   \begin{tabular}{c|cccc}
    $M_{GMO} / M_B$ & 
    Source & 
    $D$ & 
    $F$ & 
    $C$ \\
    \hline
    \hline
    $0.79 \%$ &
    ChPT &
    $0.61$ &
    $0.40$ &
    $1.2$ \\
     $1.12 \%$ &
    Lattice QCD~\cite{Lin:2007ap,Alexandrou:2006mc} &
    $0.72$ &
    $0.45$ &
    $1.6$ \\
     $1.29 \%$ &
    SU(6) &
    $3/4$ &
    $1/2$ &
    $3/2$ \\
    \hline
    \hline
       \end{tabular}
  \label{t:GMO}
  \end{center}
 \end{table}
%

The baryon Gell-Mann--Okubo mass relation is remarkably well satisfied, 
and remarkably well accounted for in SU(3) HBChPT. 
What is even more remarkable is that none of the individual baryon masses
appears to be under perturbative control in the three-flavor expansion. 
Using the ChPT estimate for the axial couplings shown in Table~\ref{t:GMO}, 
we can evaluate the NLO loop contributions to the octet baryon masses 
numerically. 
For the decuplet resonance contributions, 
we perform a subtraction so that the chiral limit mass is not renormalized:
$\mathcal{F}(m_\phi, \Delta, \mu) \longrightarrow
\mathcal{F}(m_\phi, \Delta, \mu) - \mathcal{F}(0, \Delta, \mu)$. 
The loop contributions should be of natural size at a scale 
$\mu = \Lambda_\chi$. 
Instead,
we find
$\delta M_N(\mu = \Lambda_\chi) / M_N = - 39 \%$
for the nucleon, 
$\delta M_\Lambda (\mu = \Lambda_\chi) / M_\Lambda = - 67 \%$
for the lambda, 
$\delta M_\Sigma(\mu = \Lambda_\chi) / M_\Sigma = - 89 \%$
for the sigma, 
and
$\delta M_\Xi(\mu = \Lambda_\chi) / M_\Xi = - 98 \%$
for the cascade. 
Requiring these loop contributions to be balanced by local terms is rather precarious,
and contrary to a well behaved effective theory. 
The baryon masses are just one example hinting at the ill-fated nature of SU(3) ChPT. 
Most baryon observables receive large loop contributions from kaons and etas, 
bringing the chiral expansion into question. 
The behavior, moreover, is worse for hadrons with increasing strangeness, 
as the masses exemplify.

\section{Two-Flavor Chiral Expansion}

\subsection{Schematic Example}

As the kaon and eta loops are the culprit for numerically large
contributions to the baryon masses in SU(3), 
let us first consider a schematic example of this problem. 
The mass of the sigma receives kaon contributions up to NLO
which have the form
\begin{equation} \label{eq:MX}
M_\Sigma 
= 
M^{SU(3)}
+ 
a m_K^2 
+ 
b m_K^3
.\end{equation}
The parameter 
$M^{SU(3)}$ 
is the average octet baryon mass in the SU(3) chiral limit. 
The analytic contribution 
$\propto m_K^2$ 
arises from the LO chiral symmetry breaking operators, while the non-analytic contribution 
$\propto m_K^3$ 
arises from the sunset diagram. 
We have omitted any pion and eta contributions in this schematic example. 
By virtue of the Gell-Mann--Oakes--Renner (GMOR) relation, 
we can write the kaon mass in the form
\begin{equation}
m_K^2 = \frac{1}{2} m_\pi^2 + \frac{1}{2} m_{\eta_s}^2
,\end{equation}
where $m_{\eta_s}$ is the mass of the quark basis $\overline{s}s$ meson. 
Using LO ChPT and the masses of the neutral pion and kaons, 
we have 
$m_{\eta_s}  = 0.69 \, \texttt{GeV}$, 
and a natural expansion suggests itself:
expand in powers of 
$m_\pi / m_{\eta_s} \sim 0.2$. 
This is equivalent to treating 
$m_u, m_d \ll m_s \sim \Lambda_{QCD}$. 
Carrying out this expansion on Eq.~(\ref{eq:MX}), 
we arrive at
\begin{equation} \label{eq:MX2}
M_\Sigma 
= 
M^{SU(3)}
+ 
a' m^2_{\eta_s}
+ 
a'' m_\pi^2 
+ 
b' m_{\eta_s}^3
+
b'' m_{\eta_s} m_\pi^2
+
b''' \frac{1}{m_{\eta_s}} m_\pi^4
+ 
\ldots
.\end{equation}
The omitted terms consist of higher powers of $m_\pi / m_{\eta_s}$. 
From the above form, 
the non-analytic strange quark mass dependence can be absorbed into the relevant low-energy constants
of a two-flavor chiral expansion of the sigma mass
\begin{equation}
M_\Sigma = M_\Sigma^{SU(2)} + \alpha m_\pi^2 + \beta m_\pi^3 + \ldots
.\end{equation}
The $m_\pi^3$ term did not arise from the kaon contributions considered, 
we added it for completeness.

In the two-flavor chiral expansion,
large contributions from kaons and etas
have been summed to all orders in the resulting low-energy constants. 
A well-behaved expansion in powers of 
$m_\pi / m_{\eta_s}$  
requires that thresholds for kaon production cannot be reached. 
When this condition is met, 
the kaons and eta need not appear explicitly in the effective theory, 
and their virtual loop contributions can be reordered as described here.
Such an SU(2) formulation can describe the virtual strangeness changing transitions
provided one is suitably far from these thresholds. 
We make this criterion quantitative by considering kaon production thresholds.

\subsection{Kaon Production Thresholds}

The phenomenological values for SU(3) splittings of the octet baryons are given by
\begin{eqnarray}
\delta_{N \Sigma^*} &=& 0.45 \, \texttt{GeV}, \qquad 
\delta_{\Lambda \Xi^*} = 0.42 \, \texttt{GeV}, \qquad
\delta_{\Xi \Omega} = 0.36 \, \texttt{GeV}, \qquad
\delta_{\Sigma \Xi^*} = 0.34 \, \texttt{GeV}, \nonumber \\ 
\delta_{N \Sigma} &=& 0.26 \, \texttt{GeV}, \qquad 
\delta_{\Lambda \Xi} = 0.20 \, \texttt{GeV}, \qquad
\delta_{N \Lambda} = 0.18 \, \texttt{GeV}, \qquad
\delta_{\Sigma \Xi} = 0.13 \, \texttt{GeV}
\label{eq:splits}
,\end{eqnarray}
where 
$\delta_{BB'}$ 
denotes the difference in baryon masses,  
$\delta_{BB'} = M_{B'} - M_{B}$, 
with 
$B$ 
a spin-1/2 baryon, 
and
$B'$ 
either a spin-1/2, or spin-3/2 baryon. 
While all 
$\Delta S = 1$
splittings are below threshold, 
$\delta_{BB'} < m_K$, 
with 
$m_K = 0.50 \, \texttt{GeV}$,
the spin-3/2 to spin-1/2 transitions are not considerably far from threshold. 
At first glance, 
it appears that the SU(2) theory will poorly describe the non-analyticities associated with such inelastic thresholds. 
This impression is based on the value of 
$m_K / \delta_{BB'} \sim 1$; 
which, 
however, 
is not the appropriate expansion parameter for SU(2).

To deduce the expansion parameter relevant for an SU(2) description of hyperons, 
we return to the schematic example, 
and include the SU(3) splitting, 
$\delta_{BB'}$. 
The mass of the 
$\Sigma$ 
baryon, 
for example,
receives a loop contribution from 
$K$-$N$ 
intermediate states of the form
\begin{equation} \label{eq:example}
\delta M_\Sigma
\propto
\mathcal{F} (m_K, - \delta_{N \Sigma}, \mu )
.\end{equation}
When the SU(3) splitting is ignored, 
$\delta_{N \Sigma} \to 0$, 
we recover the kaon loop contribution originally considered above,
namely
$\mathcal{F} (m_K, 0, \mu) = \pi \, m_K^3$.
For an arbitrary baryon 
$B'$, 
a $\Delta S = 1$
virtual process leads to a non-analytic contribution
of the form 
$\mathcal{F}(m_K, - \delta_{BB'}, \mu)$, 
where $B$ is the intermediate state baryon. 
Near threshold, 
$\delta_{BB'} - m_K \to 0^+$, 
this function behaves as
\begin{equation}
\mathcal{F}(m_K, - \delta_{BB'}, \mu = m_K)  \longrightarrow 2 \pi i  \, (\delta_{BB'}^2 - m_K^2 )^{3/2} + \ldots
,\end{equation}
which is dictated by the available two-body phase space at threshold, 
and the requirement that the kaon and the $B$ baryon be in a relative $p$-wave. 
Choosing the scale 
$\mu = m_K$ 
is a convenient way to remove contributions not associated with the long-distance kaon production. 
The imaginary part of 
$\delta M_{B'}$
leads to the width for 
$B' \to K \, B$
decay.

For the mass splittings listed in Eq.~(\ref{eq:splits}), 
our concern is with the region below threshold, 
$\delta_{BB'} - m_K \to 0^-$.
In this limit, 
the SU(2) treatment must fail, 
and we must address whether the physical splittings put us in this region. 
Applying the perturbative expansion about the SU(2) chiral limit 
for a generic non-analytic function $f(x)$, 
we have
\begin{equation} \label{eq:expand}
f( 2 m_K^2 - 2 \delta_{BB'}^2 )
=
f( m_{\eta_s}^2 - 2 \delta_{BB'}^2)
+
m_\pi^2 
f' (m_{\eta_s}^2 - 2 \delta_{BB'}^2 )
+
m_\pi^4
f'' (m_{\eta_s}^2 - 2 \delta_{BB'}^2 )
+ \ldots
.\end{equation}
Thus for the subthreshold case, 
the expansion parameter, 
$\epsilon_{BB'} $,
is generically of the form
\begin{equation}
\epsilon_{BB'} 
= 
\frac{m_\pi^2}{m_{\eta_s}^2 - 2 \delta_{BB'}^2}
.\end{equation}
For the strangeness transitions listed in Eq.~(\ref{eq:splits}),
we have:
$\epsilon_{N \Sigma^*} = 0.23$, 
$\epsilon_{\Lambda \Xi^*} = 0.14$, 
$\epsilon_{\Xi \Omega} = 0.09$, 
$\epsilon_{\Sigma \Xi^*} = 0.08$,
$\epsilon_{N \Sigma} = 0.05$, 
$\epsilon_{\Lambda \Xi} = 0.05$,
$\epsilon_{N \Lambda} = 0.05$,
$\epsilon_{\Sigma \Xi} = 0.04$. 
Despite the nearness of thresholds (compared to the kaon mass), 
the expansion parameters in SU(2) are all better than the 
generic expansion parameter for SU(3), 
$\epsilon \sim m_\eta / M_B = 0.5$.

%
\begin{figure}[t]
\begin{center}
\epsfig{file=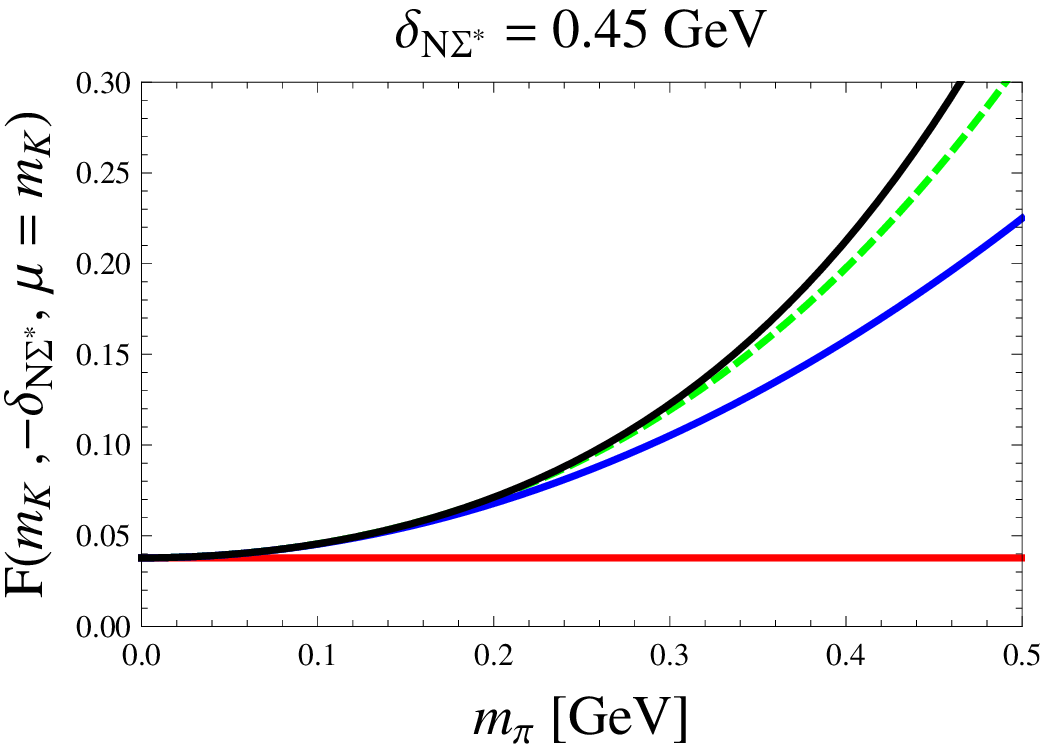,width=2.65in}
$\quad$
\epsfig{file=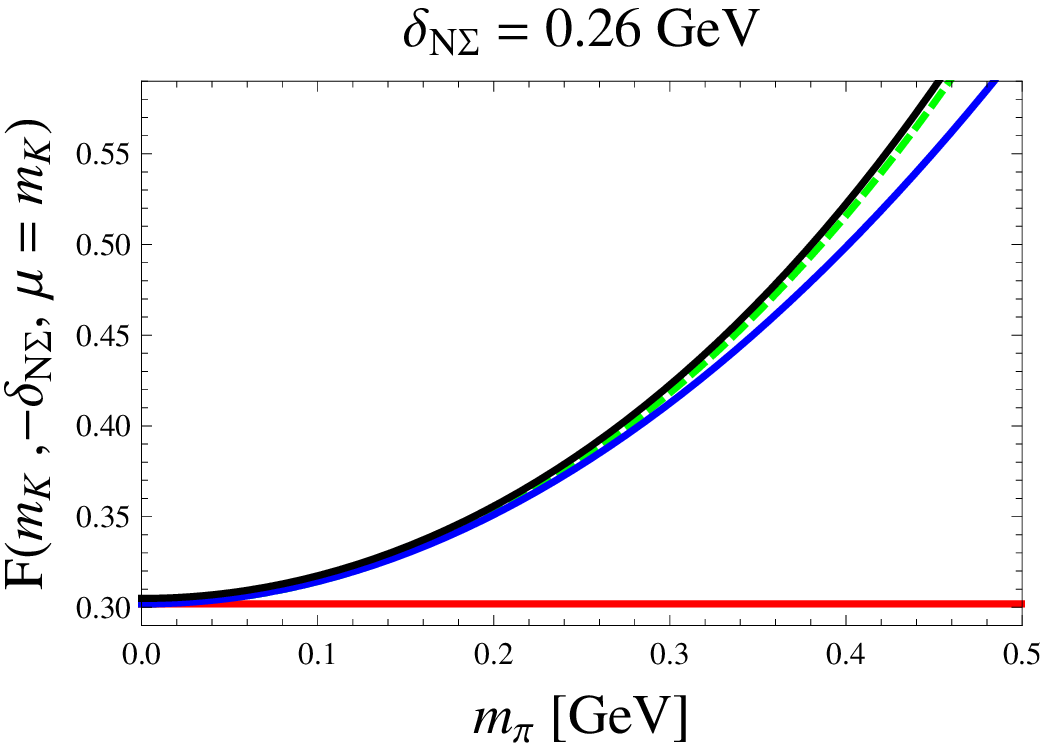,width=2.65in}
\caption{\label{f:expand} 
Virtual threshold contribution from the
$K$-$N$ 
loop diagram for the $\Sigma^*$ (left), 
and $\Sigma$ (right) baryon masses.
Plotted versus the pion mass and shown in dashed green 
is the non-analytic contribution 
$\mathcal{F}(m_K, - \delta_{BB'}, \mu = m_K)$.
Compared with this curve are three approximations
that are analytic in the pion mass squared. 
The red curve is the zeroth-order approximation, 
while the blue curve also includes the first-order correction
proportional to $m_\pi^2$, 
and finally the black curve includes all terms to 
$m_\pi^4$. 
}
\end{center}
\end{figure}
%

We can investigate the degree to which kaon thresholds 
affect hyperon masses by expanding the non-analytic 
function 
$\mathcal{F} (m_K, - \delta_{BB'}, \mu = m_K)$ 
in powers of the pion mass, 
as in Eq.~(\ref{eq:expand}).
Again we first evaluate the function at the scale 
of the kaon mass in order to remove the logs
which are not associated with the threshold, 
i.e.~the log terms have a simple series expansion in 
$m_\pi / m_{\eta_s}$, 
which is unencumbered by the threshold. 
In Figure~\ref{f:expand}, 
we show the non-analytic contribution to the 
masses of
$\Sigma^*$ and $\Sigma$ 
baryons arising from virtual 
$K$-$N$ 
fluctuations. 
This result is compared with successive approximations
derived by expanding about the SU(2) chiral limit. 
The plots show the non-analytic contribution 
associated with the virtual kaon threshold
can be captured in the two-flavor effective theory.
In SU(2), 
the kaon thresholds are described 
by a tower of terms analytic in the pion mass squared, 
but non-analytic in the strange quark mass. 
Figure~\ref{f:expand} 
confirms that the expansion in terms of 
$\epsilon_{BB'}$ 
in Eq.~(\ref{eq:expand}) is under control
for the range of values corresponding to the
$\Delta S =1$ 
transitions.
\footnote{ 
These results are encouraging, 
however, 
they cannot be definitive. 
We have estimated the mass of the 
$\eta_s$ 
meson by using the GMOR relation for the neutral kaon mass. 
Allowing the 
$\eta_s$ 
mass to vary $10 \%$
shows that one of the transitions listed has a potentially fallible expansion.  
If the mass of the 
$\eta_s$
is 
$10 \%$
smaller,
then the expansion in 
$\epsilon_{N \Sigma^*}$ 
is ill-fated. 
Beyond LO, 
we can define
$m_{\eta_s}$ 
as twice the SU(2) chiral limit value of the kaon mass
so that the expansion parameters take basically the 
same functional form, 
$\epsilon_{BB'} = (2 m_K^2 - m_{\eta_s}^2 ) / (m_{\eta_s}^2 - 2 \delta_{BB'}^2)$. 
We can then utilize SU(3) ChPT to determine the sign of the NLO correction, 
which is the sign of:
$ \frac{1}{36 \pi^2} \log \frac{m_{\eta_s} }{ \mu} + 4 \big\{ 2 L_8 (\mu) - L_5(\mu) + 2  \big[ 2 L_6(\mu) - L_4(\mu) \big] \big\}$.
Using a variety of values for the LECs determined from lattice QCD~\cite{Aubin:2004fs,Allton:2008pn}, 
and NNLO ChPT phenomenology~\cite{Amoros:2001cp}, 
we find that the net sign is positive (even when minimizing contributions from the LEC terms). 
With a positive correction to
$m_{\eta_s}$, 
the expansion parameters 
$\epsilon_{BB'}$
are all smaller than estimated, 
and the virtual
$\Sigma^* \to KN$
process is likely well described in SU(2). 
}

\subsection{Baryon Masses in SU(2) HBChPT}

%
  \begin{table}[t]
  \begin{center}
      \caption{Comparison of SU(3) ChPT and SU(2) ChPT for baryons.  
   Parameters in SU(2) are not related between the various strangeness sectors. }
   \smallskip
   \begin{tabular}{c|ccccc}
     & 
    $SU(3)$ & 
    $SU(2)_{S=0}$ &
    $SU(2)_{S=1}$ & 
    $SU(2)_{S=2}$ & 
    $SU(2)_{S=3}$ \\  
    \hline
    \hline
Expansion &
$p$
$m_\pi$
$m_K$
$m_\eta$
$\Delta$  &
$p$ 
$m_\pi$
$\Delta_{\Delta N}$ &
$p$ 
$m_\pi$
$\Delta_{\Sigma \Lambda}$
$\Delta_{\Sigma^* \Sigma}$ &
$p$ 
$m_\pi$
$\Delta_{\Xi^* \Xi}$ &
$p$ 
$m_\pi$ \\
Multiplets &
$\mathbf{8} \, B \,$ 
$\mathbf{10} \, T$ &
$\mathbf{2} \, N \,$ 
$\mathbf{4} \, \Delta$ &
$\mathbf{1} \, \Lambda \,$ 
$\mathbf{3} \, \Sigma\, $ 
$\mathbf{3} \, \Sigma^*$ &
$\mathbf{2} \, \Xi \,$
$\mathbf{2} \, \Xi^*$ &
$\mathbf{1} \, \Omega$ \\
Couplings &
$D$ 
$F$
$C$ 
$H$ &
$g_A$ 
$g_{\Delta N}$ 
$g_{\Delta \Delta}$ &
$g_{\Lambda \Sigma}$ 
$g_{\Sigma \Sigma}$
$g_{\Lambda \Sigma^*}$ 
$g_{\Sigma \Sigma^*}$ 
$g_{\Sigma^* \Sigma^*}$ &
$g_{\Xi \Xi}$
$g_{\Xi \Xi^*}$ &
\\
    \hline
    \hline
       \end{tabular}
  \label{t:compare}
  \end{center}
 \end{table}
%

\begin{figure}[t]
\begin{center}
\epsfig{file=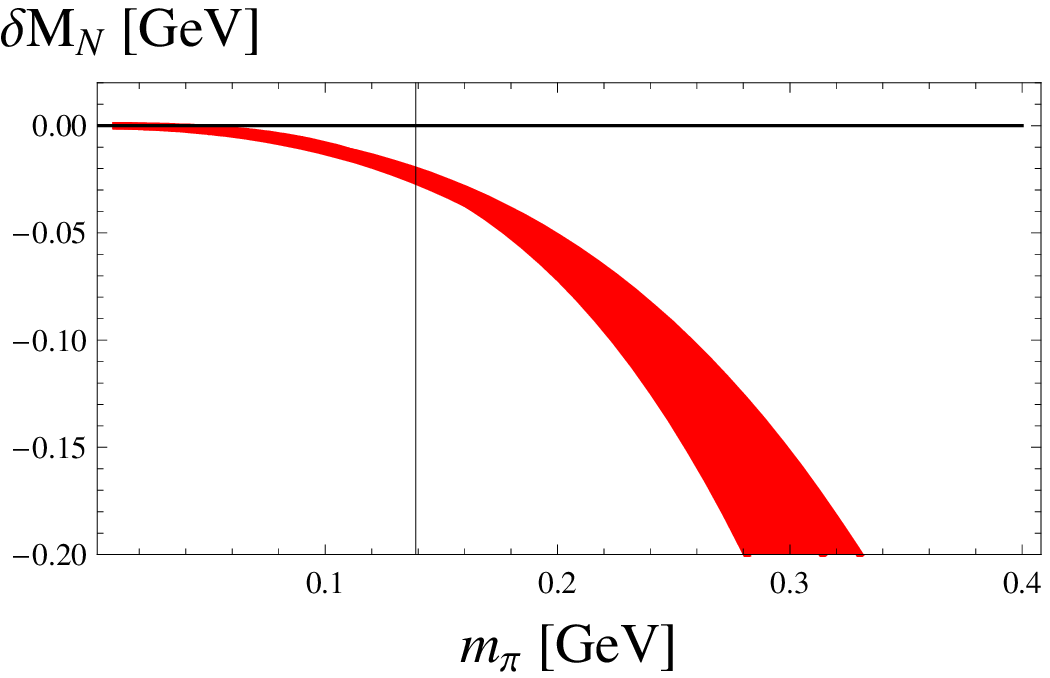,width=2.65in}
$\quad$
\epsfig{file=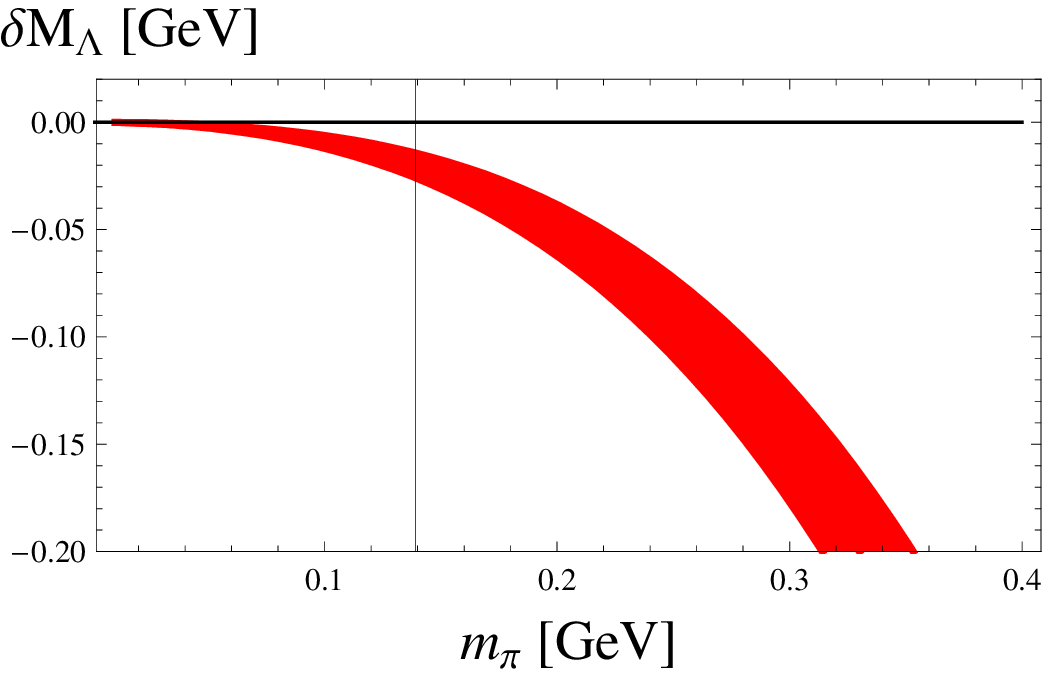,width=2.65in}
\\
\epsfig{file=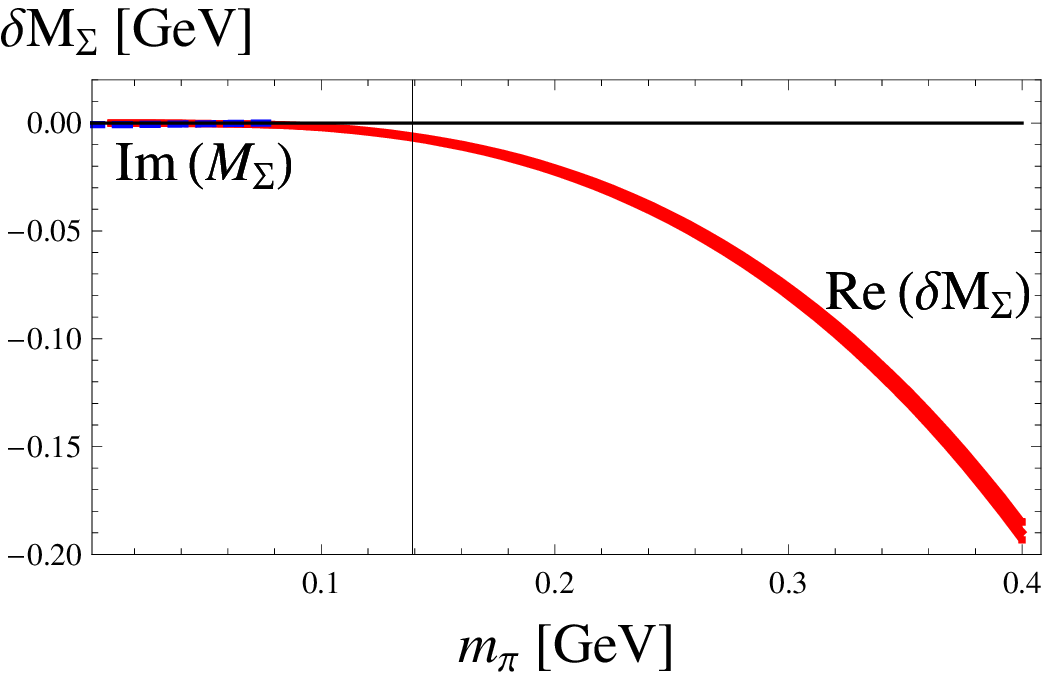,width=2.65in}
$\quad$
\epsfig{file=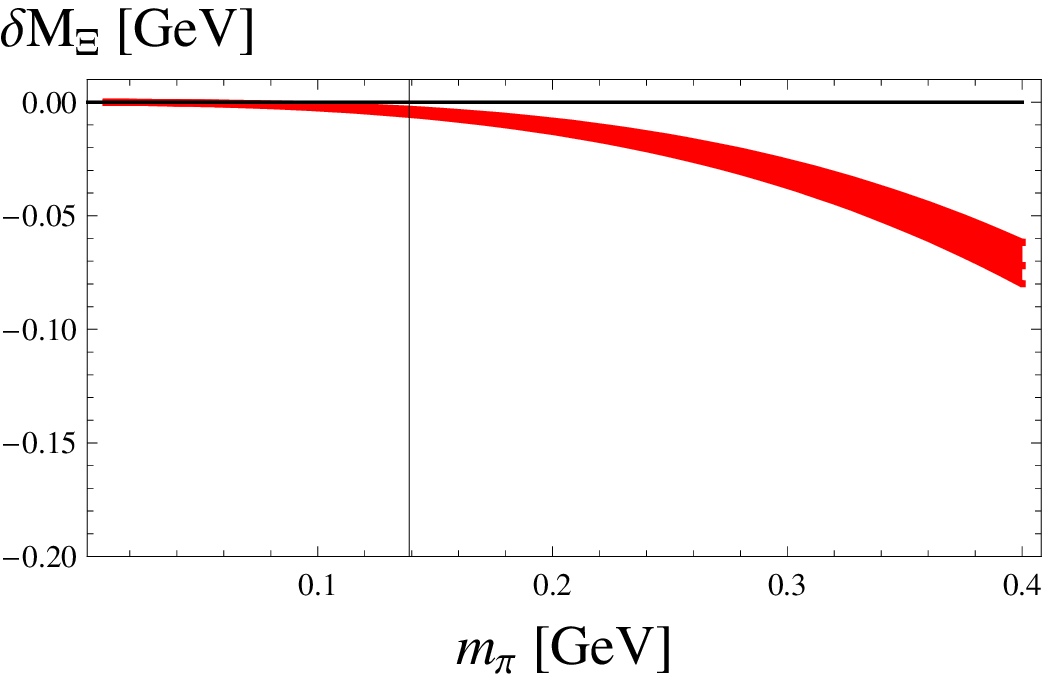,width=2.65in}
\caption{\label{f:mass} 
Behavior of the NLO contributions to the baryons masses in SU(2) HBChPT.
The bands arise from varying the renormalization scale $\mu$ about $\mu = \Lambda_\chi$.}
\end{center}
\end{figure}

Having discussed aspects of the formulation of two-flavor ChPT for hyperons, 
we turn our attention to using this theory to compute hyperon properties, 
and assess the convergence of SU(2) relative to SU(3). 
A comparison of the ingredients of SU(3) and SU(2) ChPT is presented in Table~\ref{t:compare}. 
The computation of baryon masses in SU(2) HBChPT has been carried out~\cite{Tiburzi:2008bk}.
In Figure~\ref{f:mass}, 
we show the pion mass dependence of the NLO computation of baryon masses in SU(2). 
There is marked improvement over the SU(3) chiral expansion. 
In particular the behavior of the NLO contributions is perturbative at the chiral symmetry breaking scale for a range of pion masses. 
Furthermore, 
the behavior of the SU(2) chiral expansion improves with increasing strangeness. 
This feature owes itself to two facts.
Firstly, the non-relativistic expansion improves with increasing strangeness
because the relevant expansion parameter, 
$m_\pi / M_S$, 
decreases.
Secondly, 
the pion loop contributions are smaller with increasing strangeness due to 
reduced axial couplings:
$g_A = 1.25$, 
$g_{\Sigma \Sigma} = 0.78$, 
$g_{\Xi \Xi} = 0.24$, 
and
$g_{\Delta N} = 1.48$, 
$g_{\Lambda \Sigma^*} = 0.91$,
$g_{\Sigma \Sigma^*} = 0.76$, 
$g_{ \Xi \Xi^*} = 0.69$.   
The only exception is the lambda-sigma axial coupling
$g_{\Lambda \Sigma} = 1.47$, 
although our normalization of this coupling is based on SU(3).

%
%

\section{Baryon Axial Charges}

Consideration of the baryon axial charges is analogous to the baryon masses. 
First let us consider the case of SU(3) HBChPT. 
One can derive Gell-Mann--Okubo type relations for the axial charges~\cite{Jiang:2008aqa}. 
Considering the 
$\Delta I = 1$, 
and 
$\Delta S = 1$ 
axial transition matrix elements, there are a total of 8 axial charges. 
At LO in the three-flavor expansion, 
there are only two axial couplings, $D$ and $F$. 
Thus there are six relations between the axial charges at LO. 
At NLO, 
one must consider operators with one insertion of the quark mass matrix. 
There are a total of 6 NLO operators, and consequently two non-trivial 
combinations of axial couplings that are independent of local contributions
\begin{eqnarray}
\Delta g 
&=&
2 g_{NN} - g_{N \Lambda} - g_{N \Sigma} - g_{\Lambda \Sigma} - g_{\Sigma \Sigma} + 2 g_{\Sigma \Xi}, 
\\
\Delta G 
&=&
2 g_{NN} + 2 g_{\Xi \Xi} - 2 g_{\Lambda \Sigma} + g_{N \Sigma} + g_{\Lambda \Xi} + g_{\Sigma \Xi} - g_{N \Lambda} 
.\end{eqnarray} 
These combinations can be computed from HBChPT at NLO, 
for which we find
$\Delta g = -0.0035$, and $\Delta G  = -0.017$. 
Unfortunately we cannot test these predictions against experiment as some 
of these axial charges are poorly determined, or unknown at present. 
Lattice QCD calculations, 
however, 
will be able to help us address whether SU(3) HBChPT is under control
for these combinations of axial couplings.

\begin{figure}
\begin{center}
\epsfig{file=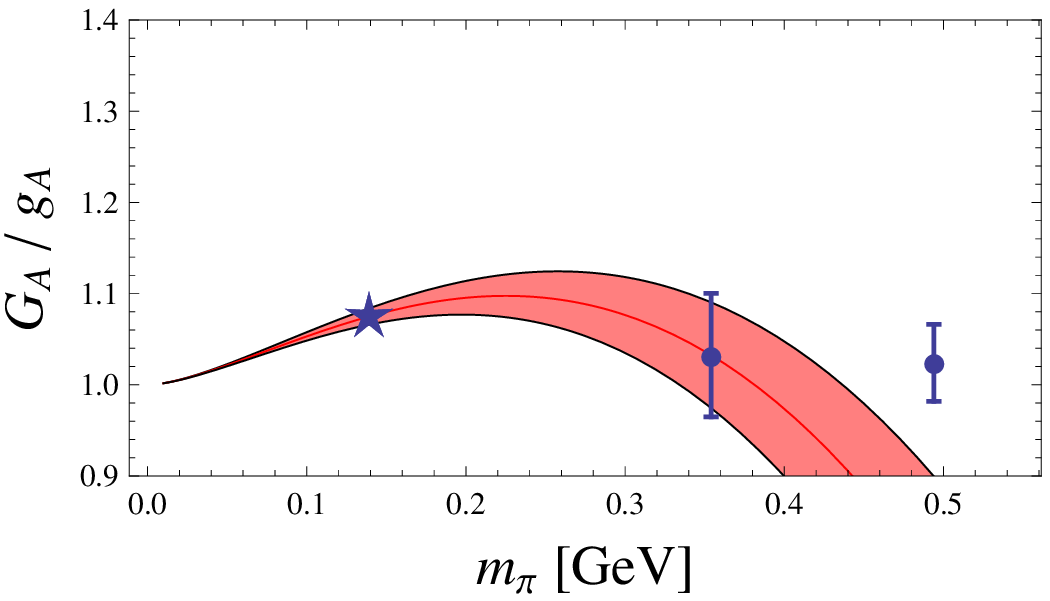,width=2.65in}
$\quad$
\epsfig{file=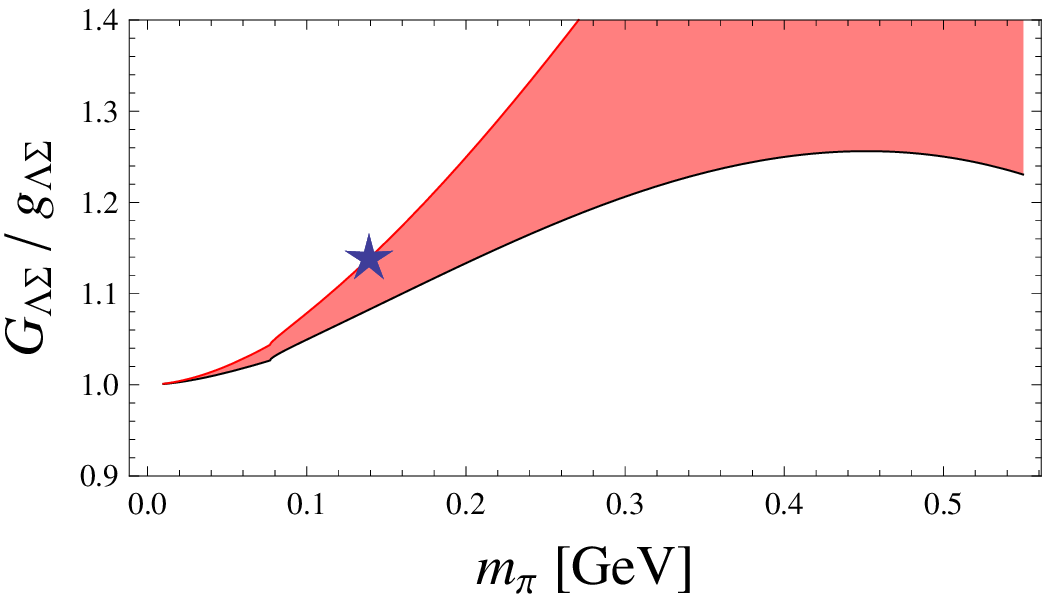,width=2.65in}
\\
\epsfig{file=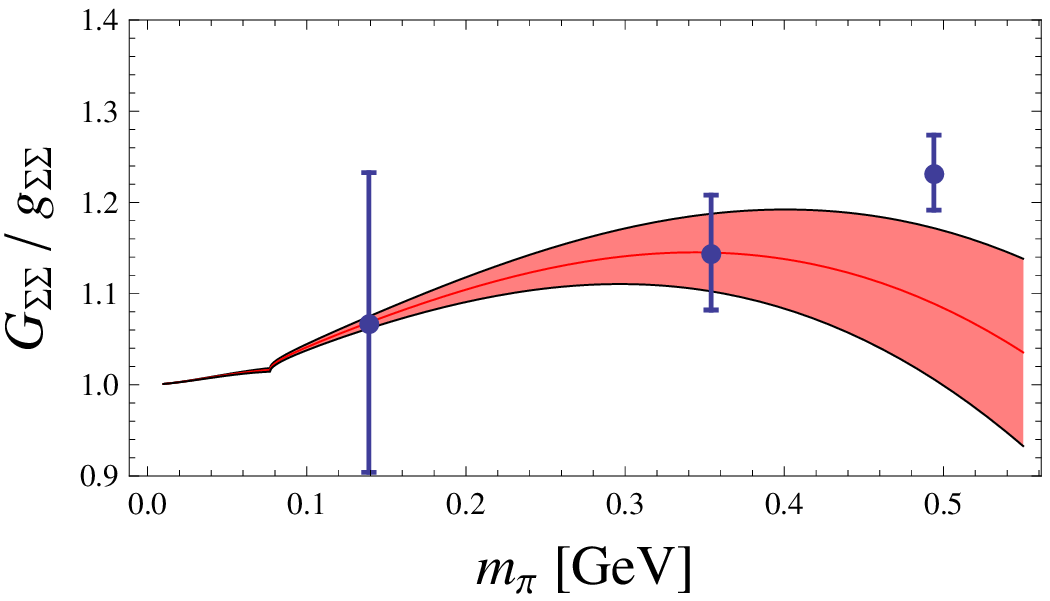,width=2.65in}
$\quad$
\epsfig{file=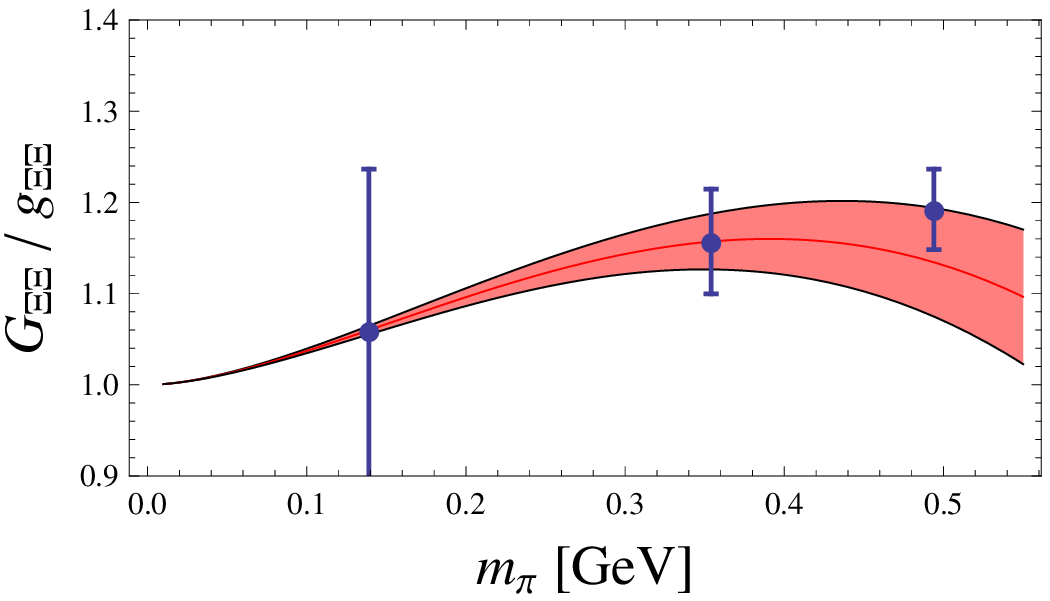,width=2.65in}
\caption{\label{f:axial} 
Pion mass dependence of the baryon axial charges in SU(2) HBChPT. 
The bands arise from varying the local contribution arising at NNLO---except for
$G_{\Lambda \Sigma}$. 
For this case, we vary the renormalization scale $\mu$ about $\Lambda_\chi$
as no lattice data exist from which to determine the NLO term.
Upper-case G's denote the axial couplings, while lower-case g's denote their chiral limit values.
Lattice data and extrapolated values are taken from~\cite{Lin:2007ap}.}
\end{center}
\end{figure}

As with the baryon masses, 
the loop contributions to individual baryon axial charges are not small in SU(3) HBChPT. 
One can use SU(2) HBChPT to compute the axial charges of hyperons, 
and this has been done for the $\Delta I = 1$ axial charges~\cite{Jiang:2009sf}. 
Results of these computations are shown in Figure~\ref{f:axial}. 
Here the pion mass dependence of the axial couplings is plotted. 
Lattice QCD data~\cite{Lin:2007ap} has been utilized to determine
the values of NLO local contributions in each strangeness sector. 
These terms make contributions to the axial charge
$G_{BB}$ 
of the form 
$A_{BB} \, m_\pi^2 / \Lambda_\chi^2$. 
We have determined
$A_{NN}(\Lambda_\chi) = -12$, 
$A_{\Sigma \Sigma}(\Lambda_\chi) = -2.9$, 
and
$A_{\Xi \Xi}(\Lambda_\chi) = - 0.22$,
for which the naturalness of these parameters increases with increasing strangeness. 
Obviously results are much better compared with SU(3) ChPT. 
These findings are promising, however, 
we must keep in mind the usual limitations of the lattice data used as input. 
Additionally due to lack of lattice QCD data, 
we are unable to include the lambda-sigma coupling in this analysis. 
Further lattice studies are needed here as the $S=1$ baryon axial charges are coupled. 
To handle the lambda-sigma mass splitting in axial current matrix elements, 
isospin twisted boundary conditions are ideal~\cite{Tiburzi:2005hg}.

%
%

\section{Summary}

The convergence of baryon chiral perturbation theory in SU(3) is precarious.
In general, observable quantities receive large contributions from kaons and etas. 
Such contributions undermine a perturbative expansion. 
Using the matching procedure between the SU(3) and SU(2) theories, 
we argued that SU(2) should exhibit better convergence
due to an expansion in 
$m_\pi / m_{\eta_s}$. 
For quantities that are far from kaon production thresholds, 
the virtual kaon and eta contributions can be reorganized into such an SU(2) chiral expansion. 
This expansion can be formulated for hyperons without explicit kaon and eta degrees 
of freedom. 
We found that even for quantum fluctuations close to kaon production thresholds, 
a new expansion parameter 
$\varepsilon_{BB'}$
underlies the two-flavor theory. 
Expansions in this quantity convert non-analytic kaon thresholds 
into a tower of analytic pion mass squared terms with 
coefficients that are non-analytic in the strange quark mass. 
Estimates of the 
$\varepsilon_{BB'}$
parameters for the hyperons show that kaon productions thresholds
will be reproduced in SU(2) perturbation theory.%
\footnote{
As we mentioned, there is possibly one exception. 
The SU(2) expansion of virtual 
$K$-$N$ 
contributions to 
$\Sigma^*$ 
properties is potentially fallible.
The sign of the NLO correction to the kaon mass in the SU(2) chiral limit, 
however, 
likely implies that the expansion parameters 
$\epsilon_{BB'}$ 
are overestimated rather than underestimated. 
}

Using the SU(2) theory of hyperons, 
we explored the chiral behavior of baryon masses and 
axial charges calculated in HBChPT. 
These explicit computations showed marked improvement over SU(3). 
The expansion, moreover, is better behaved with increasing strangeness
due to a better non-relativistic approximation, 
and axial couplings that decrease with increasing strangeness. 
Ultimately the power of SU(2) ChPT must be tested using data from lattice QCD simulations. 
Future data at light pion masses will arm us with information
about the low-energy constants in two- and three-flavor theories. 
In turn, we will be able to address the issues of convergence. 
Lattice QCD, moreover, benefits directly from SU(2) ChPT. 
At current values of the quark masses, 
what is required is pion mass extrapolation which is
efficiently handled in the two-flavor expansion. 
Finally
the combination of lattice QCD in conjunction with ChPT will enable us to
address when SU(3) ChPT is a systematic tool for the lowest-lying baryons.

\begin{acknowledgments}
The work reported here was carried out in collaboration with F.-J.~Jiang, and A.~Walker-Loud, 
with partial support from the 
U.S.~Dept.~of Energy,
Grant 
No.~DE-FG02-93ER-40762.
We gratefully acknowledge the Institute for Theoretical Physics at Bern University for their hospitality.
\end{acknowledgments}

\end{document}